\documentclass[twocolumn, times]{aastex631}
\usepackage{amsmath}
\usepackage{soul}
\defcitealias{Hankins2015}{H15}
\def\hcit{\citetalias{Hankins2015}}

\DeclareMathOperator{\erfc}{erfc}

\begin{document}

\title{High Sensitivity Beamformed Observations of the Crab Pulsar's Radio Emission}

\author[0000-0003-4530-4254]{Rebecca Lin}
\affil{Department of Astronomy and Astrophysics, University of Toronto, 50 St. George Street, Toronto, ON M5S 3H4, Canada}

\author[0000-0002-5830-8505]{Marten H. van Kerkwijk}
\affil{Department of Astronomy and Astrophysics, University of Toronto, 50 St. George Street, Toronto, ON M5S 3H4, Canada}

\correspondingauthor{Rebecca Lin}
\email{lin@astro.utoronto.ca}

\begin{abstract}
We analyzed four epochs of beamformed EVN data of the Crab Pulsar at $1658.49{\rm\;MHz}$.
With the high sensitivity resulting from resolving out the Crab Nebula, we are able to detect even the faint high-frequency components in the folded profile.
We also detect a total of $65951$ giant pulses, which we use to investigate the rates, fluence, phase, and arrival time distributions.
We find that for the main pulse component, our giant pulses represent about 80\% of the total flux.
This suggests we have a nearly complete giant pulse energy distribution, although it is not obvious how the observed distribution could be extended to cover the remaining 20\% of the flux without invoking large numbers of faint bursts for every rotation.
Looking at the difference in arrival time between subsequent bursts in single rotations, we confirm that the likelihood of finding giant pulses close to each other is increased beyond that expected for randomly occurring bursts -- some giant pulses consist of causally related microbursts, with typical separations of $\sim\!30{\rm\;\mu s}$ -- but also find evidence that at separations $\gtrsim\!100{\rm\;\mu s}$ the likelihood of finding another giant pulse is suppressed.
In addition, our high sensitivity enabled us to detect weak echo features in the brightest pulses (at $\sim\!0.4\%$ of the peak giant pulse flux), which are delayed by up to $\sim\!300{\rm\;\mu s}$.
\end{abstract}


\section{Introduction} \label{sec:intro}

The young Crab Pulsar, located within the Crab Nebula, is one of the most well-observed and studied pulsars (for a review, see \citealt{Eilek2016}).
It is visible throughout the electromagnetic spectrum, and shows many pulse components, some visible at all wavelengths, others only at specific ones.
Almost certainly, different emission locations and processes are involved, making the Crab Pulsar a unique laboratory for studying pulsar emission mechanisms.

In the radio, the mean pulse profile of the pulsar has eight components \citep{Karuppusamy2012, Hankins2015}: the low frequency component (LFC), the main pulse precursor (MP precursor), the main pulse (MP), the interpulse precursor (IP precursor), the interpulse (IP), the high frequency interpulse (HFIP), and high frequency components 1 and 2 (HFC1, HFC2, respectively).
Not all components are visible throughout the radio spectrum and some components such as the HFC1 and 2 show frequency evolution, moving in pulse phase.

At frequencies below $4{\rm\;GHz}$, the MP and IP components dominate and are particularly interesting as within their phase windows most if not all of the emission is in the form of ``giant pulses'' (GPs): narrow and bright pulses, lasting just a few microseconds and reaching fluences greater than $100{\rm\;kJy\,\mu s}$ \citep{Majid2011, Bera2019}.
Some GPs consist of more than one microburst \citep{Sallmen1999, Hankins2007, Lin2023}, with each microburst comprised of numerous nanoshots that have durations of the order of nanoseconds (e.g. \citealt{Hankins2003,Hankins2007}).

The mechanism behind GPs is not fully understood.
Apart from the Crab Pulsar, only a select group of pulsars emits GPs \citep{Johnston2004, Knight2006, Abbate2020}.
Empirically, the group has strong magnetic fields at the light cylinders, and shows non-thermal, high-energy emission at the phases GPs occur (which in the Crab Pulsar is (slightly) enhanced in rotations in which GPs are present).
This suggests that the GPs and high energy emission arise from the same emission regions, near the light cylinder.
Indeed, recent models suggest that GPs may arise from just beyond the pulsar's light cylinder, in reconnection events in the magnetospheric current sheet \citep{Philippov2018,Philippov2019}.

\begin{deluxetable*}{clclcccccc}[ht!]
\tabletypesize{\small}
\setlength\tabcolsep{4.1pt}
\tablecaption{Observation and Giant Pulse Log. \label{table:log}}
\tablenum{1}
\tablehead{\colhead{Observation}&
  &
  \colhead{$t_{\text{exp}}$ \tablenotemark{a}}&
  &
  \colhead{DM \tablenotemark{c}}&
  \multicolumn{5}{c}{\dotfill Giant Pulses \tablenotemark{d}\dotfill}\\[-.7em]
  \colhead{code}&
  \colhead{Date}&
  \colhead{(h)}&
  \colhead{Telescopes used \tablenotemark{b}}&
  \colhead{(${\rm pc\,cm^{-3}}$)}&
  \colhead{$N$}&
  \colhead{$N_{\text{MP}}$}&
  \colhead{$N_{\text{IP}}$}&
  \colhead{$r_{\text{MP}}$ (${\rm s^{-1}}$)}&
  \colhead{$r_{\text{IP}}$ (${\rm s^{-1}}$)}}
\startdata
EK036 A & 2015 Oct 18-19    & 3.27 & Ef, Bd, Hh, Jb, O8, Sv, Wb, Zc     & 56.7772 & 24553 & 21038 & 3515 & 1.786(12)           & 0.298(5)\phantom{0} \\
EK036 B & 2016 Oct 31-Nov 1 & 1.65 & Ef, Bd, Hh, O8, Sv, Wb, Zc         & 56.7668 & 19442 & 15484 & 3958 & 2.61(2)\phantom{00} & 0.668(11) \\
EK036 C & 2017 Feb 25       & 1.15 & Ef, Bd, Hh, Jb, O8, Sv, Wb, Zc     & 56.7725 &  8887 &  7532 & 1355 & 1.82(2)\phantom{00} & 0.328(9)\phantom{0}\\
EK036 D & 2017 May 28       & 1.25 & EF, Bd, Hh, Jb-II, O8, Sv, Wb, Zc  & 56.7851 & 13069 & 10797 & 2272 & 2.40(2)\phantom{00} & 0.506(11)
\enddata
\tablenotetext{a}{Total on-source exposure time, i.e., excluding telescope setup and calibration.}
\tablenotetext{b}{Abbreviations are: Ef: the~$100{\rm\;m}$ Effelsberg telescope; Bd: the~$32{\rm\;m}$ at Badary; Hh: the~$26{\rm\;m}$ in Hartebeesthoek; Jb: the~$76{\rm\;m}$ Lovell telescope; Jb-II: the~$25{\rm\;m}$ Mark II Telescope at the Jodrell Bank Observatory; O8: the~$25{\rm\;m}$ at Onsala; Sv: the~$32{\rm\;m}$ at Svetloe; Wb: the~$25{\rm\;m}$ RT1 telescope at Westerbork; and Zc: the~$32{\rm\;m}$ at Zelenchukskaya.
Other telescopes participated in some of these observation runs, but we did not use their data because of a variety of problems.}
\tablenotetext{c}{Inferred from the GPs (see \citealt{Lin2023}).}
\tablenotetext{d}{The total number and occurrence rate of GPs detected using the search algorithm described in Section~\ref{subsec:gpsearch}.
The rate uncertainty estimates, given as errors on the last digit in parentheses, assume Poisson statistics (i.e., $\sigma_r=\sqrt{N}/t_{\rm exp}$).}
\vspace{-4mm}
\end{deluxetable*}

\vspace{-4.4mm}
Investigation of the emission from the Crab Pulsar is complicated by propagation effects along the line of sight, especially at lower frequencies, $\lesssim\!2{\rm\;GHz}$.
While dispersion can be removed using coherent de-dispersion (either during recording, or afterwards with baseband data), scattering effects are difficult to remove.
This includes echoes due to propagation in the Crab Nebula itself, which sometimes are bright and obvious \citep{Backer2000, Lyne2001}, but can also be quite faint \citep{Driessen2019}, making it difficult to disentangle them from microbursts without having a good pulse sample to look for repeating structure.

Another complication in studying the emission of the Crab Pulsar is the radio-bright nebula in which the pulsar resides.
This contributes noise and hence many previous studies relied on long integrations to observe both the weaker pulse components and echoes in the average profile.
But the contribution to the noise can be reduced by resolving the nebula, using large dishes or arrays, such as the VLA, Arecibo, and Westerbork \citep{Moffett1996, Cordes2004, Karuppusamy2010, Lewandowska2022}.

In this paper, we use the European VLBI Network (EVN) to resolve out the Crab Nebula and obtain high sensitivity data.
In Section~\ref{sec:obs}, we describe our observations and data reduction, and in Section~\ref{sec:pulse_profile}, we present the resulting pulse profiles and the components that are detectable at our high sensitivity.
We turn to an analysis of GPs in Section~\ref{sec:gps}, investigating their rates, fluence, phase, and arrival time distributions, as well as weak echoes seen in the brightest GPs.
We summarize our findings in Section~\ref{sec:conclusion}.

\section{Observations and Data Reduction} \label{sec:obs}

We analyze observations of the Crab Pulsar taken by the EVN, projects EK036~A-D, at four epochs between 2015 Oct and 2017 May (see Table~\ref{table:log}).
Throughout these observations, calibrator sources were also observed resulting in breaks in our data.
While many dishes participated in these observations, for our analysis we only use telescope data that had relatively clean signals across the frequency range of $1594.49\!-\!1722.49{\rm\;MHz}$ in both circular polarizations.
At each single dish, real-sampled data were recorded in either 2 bit MARK 5B or VDIF format\footnote{For specifications of MARK5B and VDIF, see \url{https://www.haystack.mit.edu/haystack-memo-series/mark-5-memos/} and \url{https://vlbi.org/wp-content/uploads/2019/03/VDIF_specification_Release_1.1.1.pdf}, respectively.}, covering the frequency range in either eight contiguous $16{\rm\;MHz}$ wide bands or four contiguous $32{\rm\;MHz}$ wide bands.

For these datasets, single dish data were processed and then combined coherently to form a tied-array beam as described in \citet{Lin2023}.
The resulting RFI-removed, normalized, de-dispersed (using dispersion measures (DMs) listed in Table~\ref{table:log}), parallactic angle corrected, and phased baseband data were squared to form intensity data.
As in \citet{Lin2023}, we estimate the system equivalent flux density (SEFD) for the phased EVN array as $(S_{\text{CN}} + \langle S_{\text{tel}}\rangle)/N_{\rm tel}\approx\!140\!-\!160{\rm\;Jy}$, where $S_{\text{CN}}\approx833{\rm\;Jy}$ is the SEFD of the Crab Nebula at our observing frequency \citep{Bietenholz1997}, $\langle S_{\text{tel}}\rangle\simeq300{\rm\;Jy}$ is the average nominal SEFD of the telescopes\footnote{\url{http://old.evlbi.org/cgi-bin/EVNcalc}.} and $N_{\rm tel}=7$ or $8$ is the number of telescopes used.
By combining the single dishes into a synthesized beam, we resolve out the radio-bright Crab Nebula and increase our sensitivity, thus allowing us to investigate the weaker radio emission of the Crab Pulsar.

\begin{figure*}[ht!]
    \centering
    \includegraphics[width=0.985\textwidth,trim=0 0 0 0,clip]{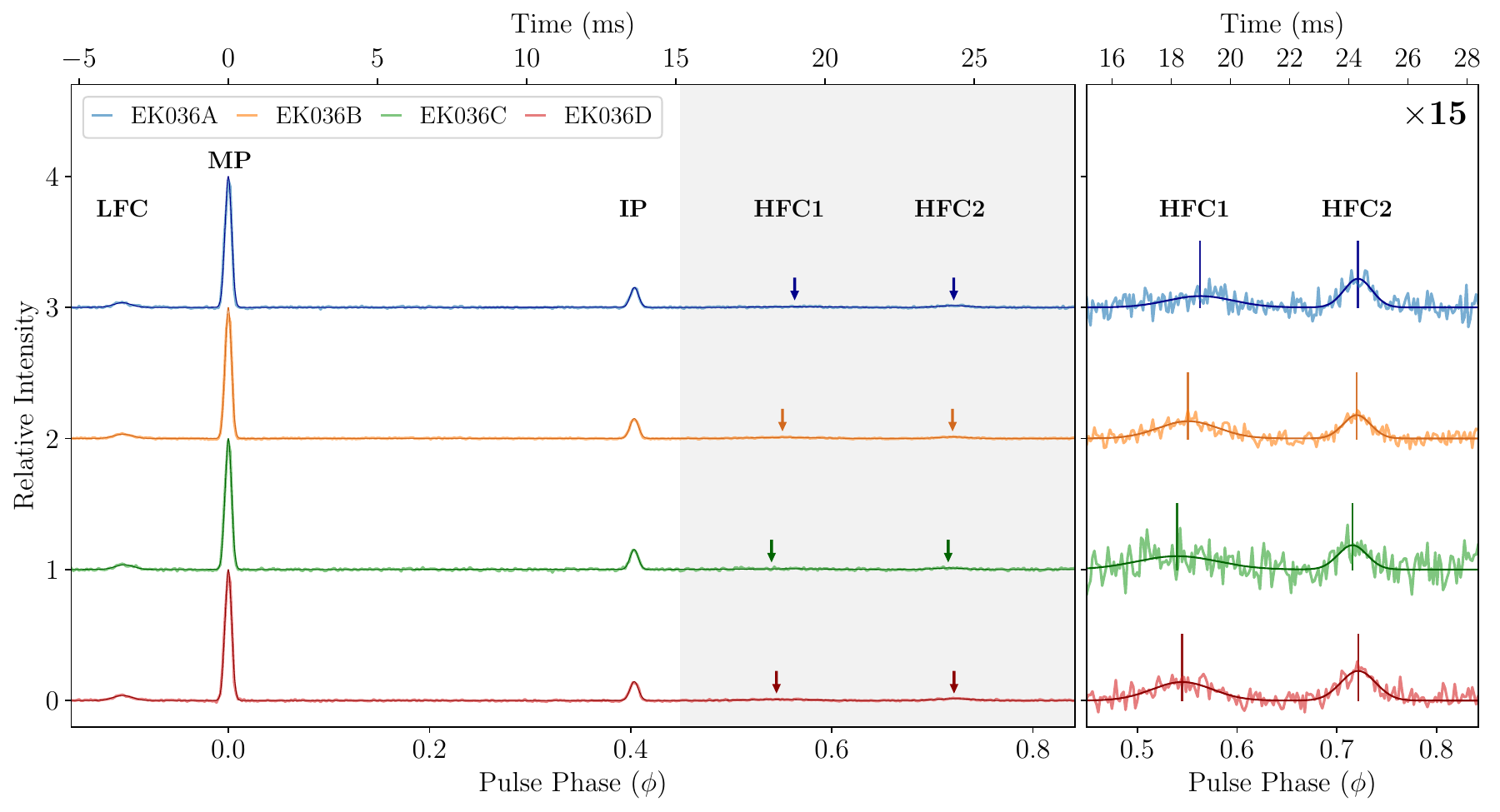}
    \caption{Folded pulse profile of the Crab Pulsar at $1658.49{\rm\;MHz}$ from EK036 observations in $512$ phase bins centered on the MP.
    At this frequency, 5 components: LFC, MP, IP, HFC1 and HFC2 are visible.
    In the left panel, the profiles are normalized to their peak MP component.
    As the HFC1 and HFC2 components (indicated by arrows) are very faint, we show the grey region of the left panel zoomed in by a factor of $15$ in the right panel, with vertical lines marking the peak of these components.}
    \label{fig:pulseprofile_EVN}
    \vspace{4mm}
\end{figure*}

\section{Pulse Profiles}\label{sec:pulse_profile}

For each of the phased EVN datasets, we create folded pulse profiles using polyco files generated with {\sc tempo2} \citep{Hobbs2012} from the monthly Jodrell Bank Crab Pulsar ephemerides\footnote{\url{http://www.jb.man.ac.uk/~pulsar/crab.html}.} \citep{Lyne1993} and DM from Table~\ref{table:log}.
We averaged over all frequencies and used $512$ phase bins, rotating in phase such that the MP is at phase $0$.
We show the resulting profiles in Figure~\ref{fig:pulseprofile_EVN}, with
each profile scaled to its maximum to ease comparison.
With our high sensitivity, we can see all five pulse components expected from the multifrequency overview of \citet{Hankins2015}, corresponding to the LFC, MP, IP, HFC1 and HFC2 (with the latter two detected at $\sim\!1.66{\rm\;GHz}$ for the first time).

We fit the pulse components in the EK036 datasets with five Gaussians to look for possible changes, both between our epochs and relative to the compilation from \citet{Hankins2015}.
Our fitted parameters are presented in Table~\ref{table:pulseprofile_fit}, together with the values inferred from \citet{Hankins2015}.
One sees that the results for our four observations are all consistent.
At $1.4{\rm\;GHz}$, \citet{Lyne2013} found that the separations between the MP and IP and between the MP and LFC increase at a rate of $0\fdg5\pm0\fdg2$ per century and $11\arcdeg\pm2\arcdeg$ per century, respectively.
Using these rates, we expect pulse phase changes for the IP and LFC of $\sim\!0\fdg008$ and $\sim\!0\fdg17$, respectively, which are not detectable within our uncertainties.

\begin{deluxetable}{lllll}
\tabletypesize{\small}
\setlength\tabcolsep{5.7pt}
\tablecaption{Properties of the Pulse Profile Components.\label{table:pulseprofile_fit}}
\tablenum{2}

\tablehead{\colhead{Pulse}&
  \colhead{Obs./}&
  \colhead{Amplitude}&
  \colhead{Pulse Phase}&
  \colhead{FWHM}\\[-.7em]
  \colhead{Comp.}&
  \colhead{Ref.}&
  \colhead{(\%)}&
  \colhead{(deg.)}&
  \colhead{(deg.)}}
\startdata
LFC\dotfill
& A     &\phn 3.6(3)   & $-38.0(3)$   &\phn 7.5(6)       \\
& B     &\phn 3.35(17) & $-37.67(19)$ &\phn 7.7(4)       \\
& C     &\phn 3.7(2)   & $-37.2(3)$   &\phn 7.7(6)       \\
& D     &\phn 3.9(2)   & $-37.8(2)$   &\phn 8.1(5)       \\
& \hcit & \nodata      & $-35.78(14)$ &\phn 7.2(12)      \\[.8ex]
MP\dotfill
& A     &              &              &\phn 2.786(11)    \\
& B     &              &              &\phn 2.708(7)\phn \\
& C     &              &              &\phn 2.756(11)    \\
& D     &              &              &\phn 2.836(9)\phn \\
& \hcit &              &              &\phn 3.9(11)      \\[.8ex]
IP\dotfill
& A     &     15.2(4)  & 145.38(4)    &\phn 3.48(10)     \\
& B     &     15.2(2)  & 145.28(3)    &\phn 3.59(7)\phn  \\
& C     &     15.3(4)  & 145.25(4)    &\phn 3.46(10)     \\
& D     &     14.4(3)  & 145.28(4)    &\phn 3.59(8)\phn  \\
& \hcit & \nodata      & 145.25(4)    &\phn 5.4(11)      \\[.8ex]
HFC1$\ldots$
& A     &\phn 0.58(13) & 203(3)       &    28(7)   \\
& B     &\phn 0.88(9)  & 198.4(13)    &    25(3)    \\
& C     &\phn 0.68(12) & 194(3)       &    34(7)   \\
& D     &\phn 0.94(11) & 196.2(15)    &    36(5)    \\
& \hcit & \nodata      & 198.2(8)     &    25(5)    \\[.8ex]
HFC2$\ldots$
& A     &\phn 1.5(2)   & 259.7(8)     &    11.8(19)    \\
& B     &\phn 1.19(14) & 259.2(7)     &    11.7(16)    \\
& C     &\phn 1.23(19) & 257.7(9)     &    12(2)\phn    \\
& D     &\phn 1.51(15) & 259.8(7)     &    14.8(16)    \\
& \hcit & \nodata      & 259.1(4)     &    11.6(12) \\
\enddata
\tablecomments{Amplitudes and phases are relative to the MP. \citetalias{Hankins2015} refers to \cite{Hankins2015}, and corresponding values are from evaluating the fits presented in his Tables 2 and 3 at our central observing frequency of $1658.49{\rm\;MHz}$.
The phases for the LFC and IP have been extrapolated to MJD 57607 (midway between EK036~A and D) using $d\phi/dt$ values from \citet{Lyne2013}.
Numbers in parentheses are $1\sigma$ uncertainties in the last digit.}
\vspace{-10.7mm}
\end{deluxetable}

\vspace{-4.3mm}
Comparing with \citet{Hankins2015}, we find good agreement in pulse phase for all components (though now we do need to take into account the drift in pulse phase).
We noticed, however, that while the widths of our LFC, HFC1 and HFC2 are consistent with those given by \citet{Hankins2015}, the widths of the MP and IP seem smaller, even if they are still within the nominal, rather large uncertainties of \citet{Hankins2015}.
Looking in more detail at their Figure~3 with measurements, one sees considerable scatter for the MP and IP, even though those strong, narrow peaks should be the easiest to measure.
This might suggest that some profiles were slightly smeared (e.g., because the data were not dedispersed to exactly the right DM, which is known to vary for the Crab Pulsar, or because of changes in scattering timescale at lower frequencies, see \citealt{McKee2018}).
For a comparison with recent data, we estimated widths from the $2\!-\!4$ and $4\!-\!6{\rm\;GHz}$ pulse profiles in Figure~1 of \citet{Lewandowska2022}, which were taken using the VLA in D configuration to resolve out the Crab Nebula and thus have high signal-to-noise ratio; we find these are all consistent with ours.

At lower frequencies, the pulse profiles often show echo features
(e.g., \citealt{Driessen2019}).
At our frequencies, those are expected to be too weak at delays where they might be seen in the folded pulse profile, and indeed we see none.
However, at frequencies like ours, echoes can still be seen in individual pulses.
For instance, at $1.4{\rm\; GHz}$, \citet{Crossley2004} saw that individual bright pulses all had an echo delayed at $\sim\!50{\rm\;\mu s}$ (which had no counterpart at $4.9{\rm\;GHz}$).
From aligning GPs before stacking them in our datasets, \citet{Lin2023} also saw hints of echo features within $\sim\!25{\rm\;\mu s}$ of the peaks of GPs in EK036~B and D.
In Section~\ref{subsec:scattering}, we confirm echoes in our data using a more careful analysis, finding that for EK036~D faint echoes are visible out to to $\sim\!300{\rm\;\mu s}$.

\section{Giant Pulses}\label{sec:gps}

\subsection{Search}\label{subsec:gpsearch}

\begin{figure}[t!]
    \centering
    \includegraphics[width=0.47\textwidth,trim=0 0 0 0,clip]{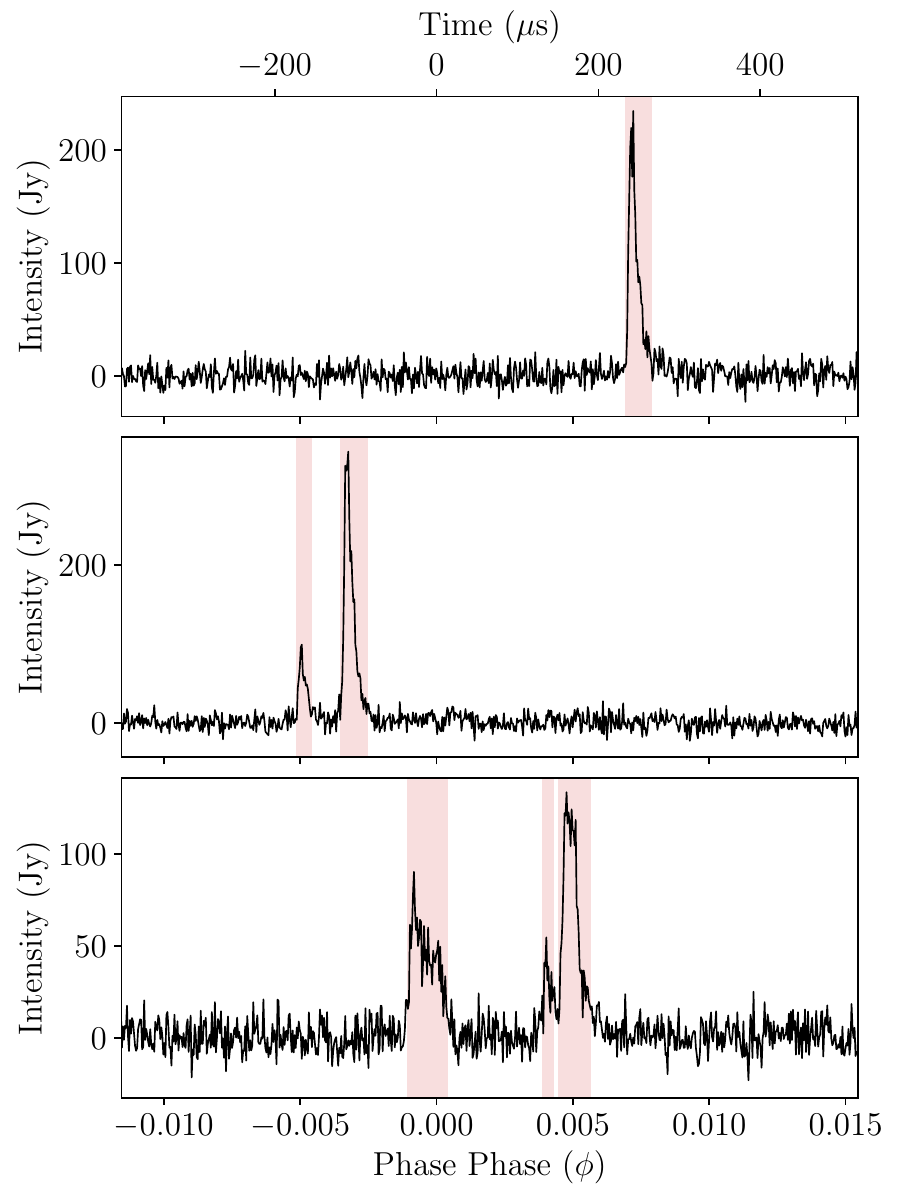}
    \caption{
    Sample MP pulse rotations with GPs as detected by our algorithm (see Section~\ref{subsec:gpsearch} for details), shown at a time resolution of $1.25{\rm\;\mu s}$.
    \emph{Top\/}: Single pulse with scattering tail.
    \emph{Middle\/}: Two pulses, each with their own scattering tail.
    \emph{Bottom\/}: A profile showing the difficulties inherent in classifying pulses: our algorithm found three pulses,
    but if another algorithm were to classify this as two or four pulses, that would also seem reasonable.
    }\label{fig:ek036d_detections}
\end{figure}
In \citet{Lin2023}, we searched for GPs by flagging peaks above $8\sigma$ in a $16{\rm\;\mu s}$ wide running average of the intensity time stream.
While we reliably found GPs, the long time window meant we could not distinguish between bursts arriving in quick succession within that time window.
Hence, the previous technique was unsuitable for one of our goals, of measuring arrival time differences between bursts, including between the microbursts that GPs sometimes are composed of.
Below, we describe a revised technique, which allows us to more reliably identify multiple bursts (see Figure~\ref{fig:ek036d_detections}).
Unsurprisingly, with our new technique we detected more multiple bursts than we had previously, as can be seen by comparing numbers listed in Section~6.3 of \citealt{Lin2023}) with those in Table~\ref{table:bursts}.

\begin{figure*}[t!]
    \centering
    \includegraphics[width=0.985\textwidth,trim=0 0 0 0,clip]{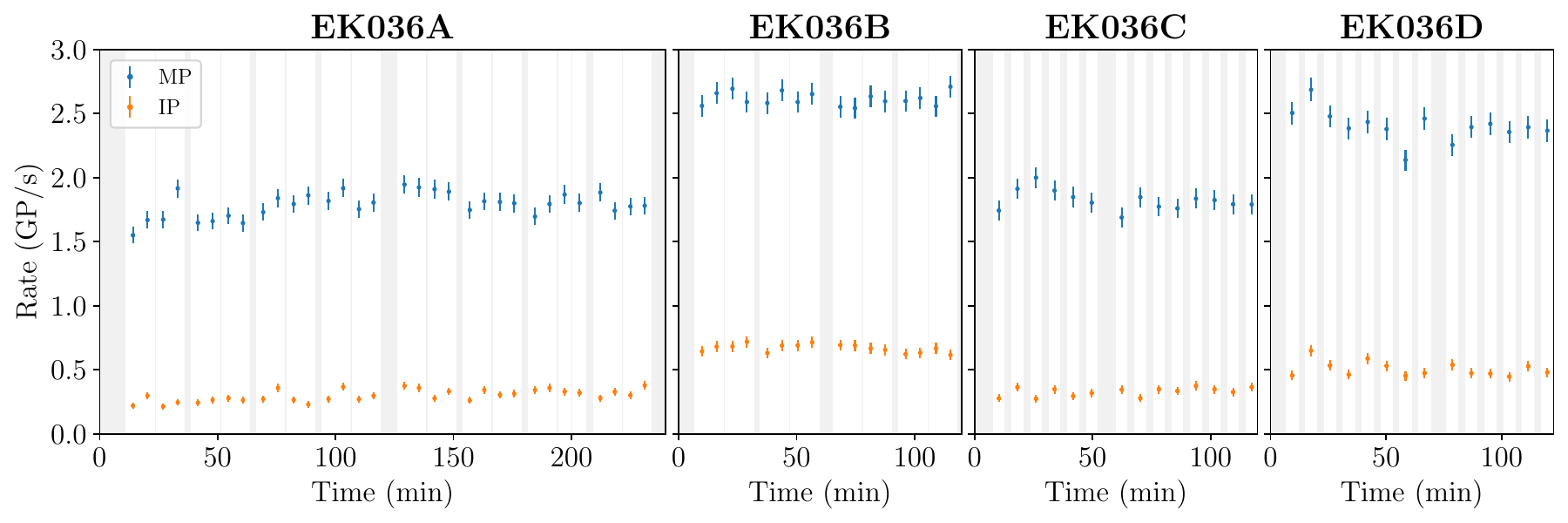}
    \caption{GP pulse detection rates in each EK036 observation.
    Times when the telescope was not observing the Crab Pulsar are shaded grey.
    The MP (blue) and IP (orange) detection rates appear to scale together and are relatively constant across each observation.
    }\label{fig:ek036_rates}
    \vspace{4mm}
\end{figure*}

For every pulsar period in the EK036 dataset, we take $2.0{\rm\;ms}$ snippets of baseband data centered at the MP and IP component phase windows (roughly $2$ times the size of the pulse component determined from the folded pulse profile) and create pulse intensity stacks for each component\footnote{We only search for GPs inside these windows since \cite{Lin2023} found none outside for the same dataset.}.
We average these stack across the eight frequency bands and bin over 10 time samples, or $0.625{\rm\;\mu s}$, a value chosen to be large enough for a reliable GP detection yet well less than the scattering timescale of $\sim\!5{\rm\;\mu s}$ during these observations \citep{Lin2023}.
To detect GPs, we first subtract the off-pulse region (determined from the $0.5{\rm\;ms}$ region on either side of each pulse stack), then filter with a uniform filter of size $5$ ($3.125{\rm\;\mu s}$), and finally record all samples above a detection threshold of~$5\sigma$.

To turn these sets of above-the-noise locations into detections of individual GPs, we use the following three-step process\footnote{Using the {\texttt binary\_closing}, {\texttt binary\_opening} and {\texttt binary\_dilation} functions, respectively, from {\sc scipy}'s multidimensional image processing functions \citep{Scipy2020}.}.
First, we connect detections within $8$ samples ($5{\rm\;\mu s}$, i.e., of order the scattering time), since those are likely related.
Second, we remove detections spanning $4$ samples ($2.5{\rm\;\mu s}$) or less, since these are likely spurious.
Third, we increase the width of a detection by $4$ samples ($2.5{\rm\;\mu s}$) on either side, mostly to ensure that if we integrate over the mask, we will capture most of the flux independent of pulse strength.
With this procedure, the minimum final pulse width is $8.125{\rm\;\mu s}$, slightly larger than the scattering timescale, and we confidently detect pulses above a threshold of $\sim\!0.15{\rm\;kJy\,\mu s}$.
The brightest GP we detect has a fluence of $\sim\!560{\rm\;kJy\,\mu s}$.
With our relatively high initial detection threshold, we do not find any GPs outside our pulse windows, suggesting that we have no false detections in our sample.
Nevertheless, as can be seen from the overall pulse statistics in Table~\ref{table:log}, we find many GPs, about $2\!-\!3$ per second or about one for every dozen pulsar rotations.

In some pulse rotations, we detect more than one distinct GP, where ``distinct'' means that the pulse is separated by at least $5{\rm\;\mu s}$ (roughly the scattering timescale) from another pulse at our detection threshold.
Here, we note that whether or not a GP is detected as single or multiple depends on the detection threshold: a GP classified as a single one at our threshold might be classified as separated at a higher threshold if it has two bright peaks with some flux in between (e.g., because the scattering tail of the first peak overlaps with the start of the next one, or a weaker burst fills in the space in between).
This dependence on detection threshold may explain why \citet{Bhat2008} found no pulses wider than $10{\rm\;\mu s}$, as they took a high detection cutoff, of $3{\rm\;kJy\,\mu s}$.
This kind of arbitrariness seems unavoidable given the variety in pulse shapes that we see; it often is a rather subjective decision on what to take as a single bursts.
To give a sense, we show in Figure~\ref{fig:ek036d_detections} an example of a pulse rotation with a single burst as well as two examples of rotations with multiple bursts.
In Section~\ref{subsec:arrival_times}, we estimate the fraction of multiple bursts that is causally related from the statistics of pulse separations.

\subsection{Rates}\label{subsec:rates}

With the high sensitivity of the phased EVN array, we detected a total of $65951$ GPs over $7.32{\rm\;hr}$, implying an average detection rate of $2.5{\rm\;s^{-1}}$.
From Table~\ref{table:log}, one sees that the rates are not the same for each epoch.
Comparable detection rates are seen for both MP and IP GPs in EK036~A and C, but those are about a factor $2$ smaller than the rates for EK036~B and D (which are comparable to each other).

\begin{figure*}[t!]
    \centering
    \includegraphics[width=0.985\textwidth,trim=0 0 0 0,clip]{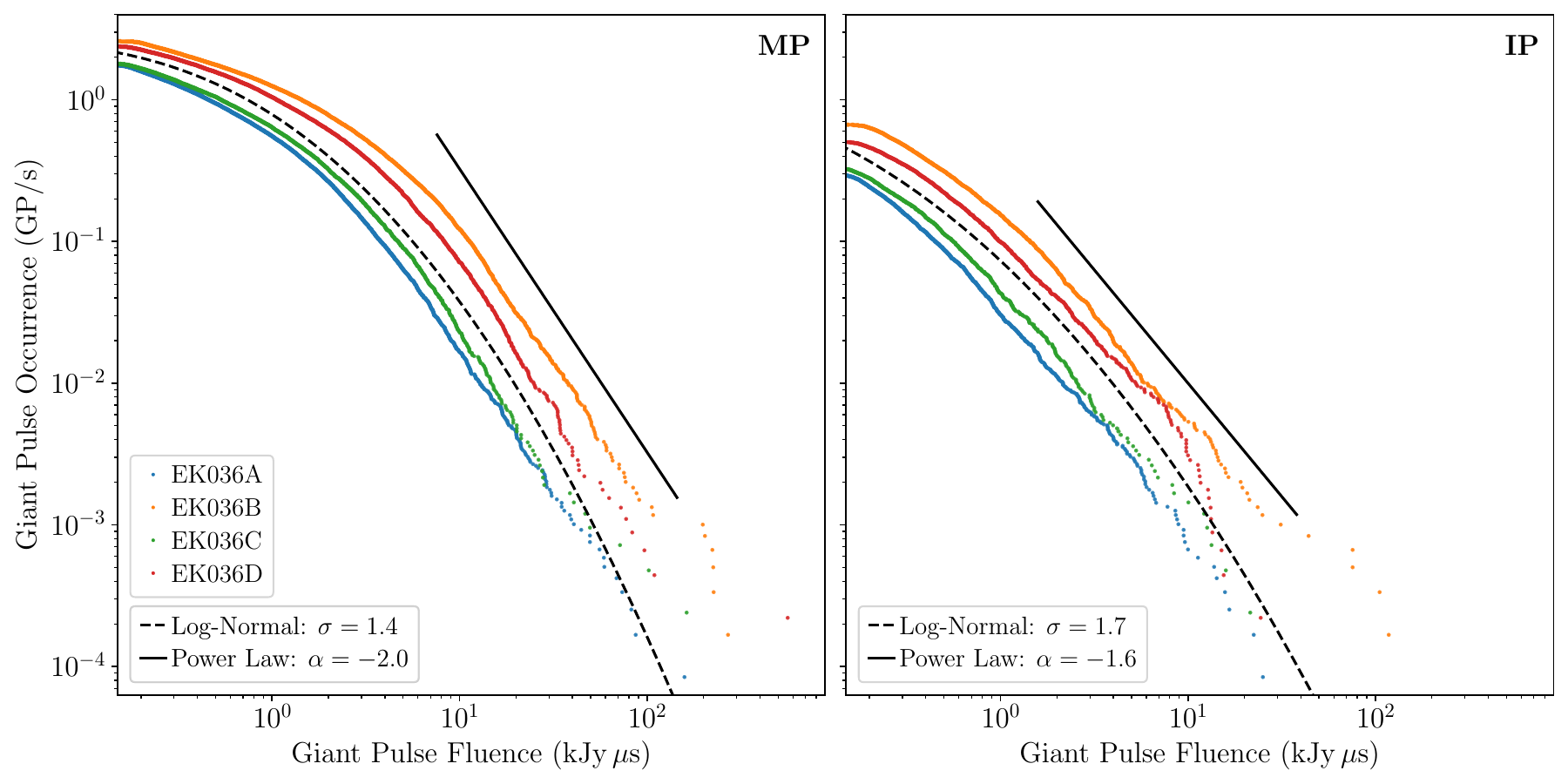}
    \caption{
    Reverse cumulative GP fluence distribution showing the occurrence rates of GPs.
    For comparison, power-law distributions (solid black lines) and log-normal distributions (dashed black line) are shown, with indices $\alpha$ and widths $\sigma$ as listed in the legend.
    }\label{fig:ek036_energydistribution}
    \vspace{4mm}
\end{figure*}

Similar changes in detection rate were found for bright pulses by \cite{Lundgren1995} at $800{\rm\;MHz}$, \cite{Bera2019} at $1330{\rm\;GHz}$ and by \cite{Kazantsev2019} at $111{\rm\;MHz}$.
\cite{Lundgren1995} suggests that almost certainly, these are due to changes in the scattering screen, which are known to cause changes in the scattering time on similar timescales and are expected to cause changes in magnification as well.
To verify that there are no variations at shorter timescales, we calculated rates at roughly $5{\rm\;min}$ intervals.
As can be seen in Figure~\ref{fig:ek036_rates}, we find that in a given epoch, the rates are indeed steady.

\subsection{Fluences}
\label{subsec:fluence}

\begin{figure*}[ht!]
    \centering
    \includegraphics[width=0.985\textwidth,trim=0 0 0 0,clip]{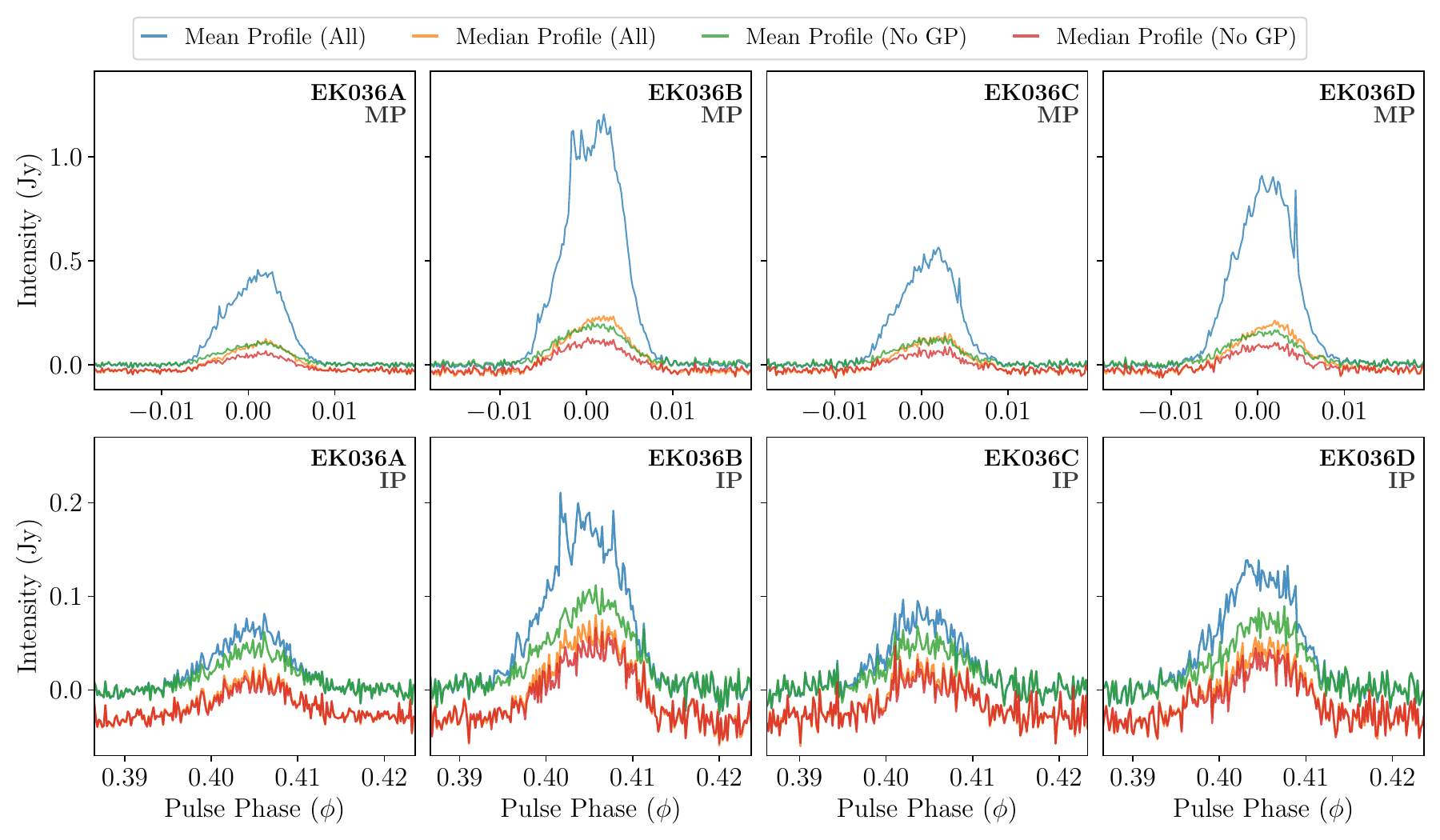}
    \caption{
    Mean and median MP and IP pulse profiles obtained using all pulse rotations (in blue and orange, respectively) and using only those in which no GPs were detected (green and red, respectively) in $6.25{\rm\;\mu s}$ bins.
    Note that because the noise in an individual profile is not normally distributed, but rather follows a $\chi_k^{2}$ distribution, the median is slightly below zero in the off-pulse region, by $(1-2/3k)^3-1\simeq-6/9k\simeq-0.0002$ of the SEFD of $\sim\!150{\rm\;Jy}$ (Section~\ref{sec:obs}), or $\sim\!-0.03{\rm\;Jy}$ given $k=3200$ degrees of freedom (complex dedispersed timestream squared, averaged over 2 polarizations, 8 bands, and 100 time bins).
    }
    \label{fig:ek036_regular_emission}
    \vspace{4mm}
\end{figure*}

The fluence distribution of the Crab Pulsar's GPs is typically described by power-law approximations to the reverse cumulative distribution,
\begin{equation}
    N_{\rm GP}(E > E_{0}) = C E_{0}^{\alpha},
\end{equation}
where $\alpha$ is the power-law index, $C$ a proportionality constant, and $E_{0}$ the GP fluence such that $N_{\rm GP}(E > E_{0})$ is the occurrence rate of GPs above $E_{0}$.
For our data, one sees in Figure~\ref{fig:ek036_energydistribution}, that for all observations the distributions indeed appear power-law like at high fluence, with $\alpha\approx\!-2.0$ and $-1.6$ for MP and IP, respectively.
These values are roughly consistent with values found at similar frequencies: e.g., \citet{Popov2007} find $-1.7$ to $-3.2$ for MP GPs and $-1.6$ for IP GPs at $1197{\rm\;MHz}$, and \citet{Majid2011} finds $\alpha=-1.9$ for the combined MP and IP distribution at $1664{\rm\;MHz}$.

However, as noted by \citet{Hankins2015} already, the power-law indices show large scatter and should be taken as roughly indicative only, showing, e.g., that at higher frequencies, very bright pulses are relatively rare.
Indeed, in our data, like in more sensitive previous studies (e.g., \citealt{Lundgren1995, Popov2007, Bhat2008, Karuppusamy2010}), the fluence distribution clearly flattens at lower fluences.
At the very low end, this is because our detection method misses more pulses, but the changes above $\sim\!0.2{\rm\,kJy\,\mu s}$ are real.
This turnover may at least partially explain why a variety of power-law indices was found previously, as the measured index will depend on what part of the fluence distribution is fit (which will depend also on the magnification by scattering), as well as why for very high fluences, well away from the turn-over, the power-law index seems fairly stable \citep{Bera2019}.

Comparing the distributions for the different epochs, one sees that they are very similar except for a shift left or right in the figure.
This confirms that the differences in rates seen between the epochs are due differences in magnification due to scintillation (and not due to the Crab Pulsar varying the rate at which pulses are emitted, which would, to first order, shift the distributions up and down).

As the fluence distributions looked roughly parabolic in log-log space, we also show  cumulative log-normal distributions in Figure~\ref{fig:ek036_energydistribution}, of the form,
\begin{equation}
  N_{\rm GP}(E > E_{0}) = \frac{A}{2}\left[\erfc\left(\frac{\ln E_0 - \mu}{\sigma\sqrt{2}}\right)\right],
\end{equation}
where $A$ is a scale factor, $\mu$ and $\sigma$ are the mean and standard deviation of $\ln E_0$, and $\erfc$ is the complementary error function.
One sees that these describe the observed cumulative distributions quite well.

If the intrinsic distributions were log-normal, it would imply that especially for the MP, most of the flux is already captured and that the total rate of GPs is not much larger than our detection rate.
For the log-normal distribution shown in Figure~\ref{fig:ek036_energydistribution}, for the MP, $A = 2.7{\rm\;s^{-1}}$ and the mean GP fluence is $\langle E\rangle = \exp(\mu + \frac{1}{2}\sigma^{2}) = 1.2{\rm\;kJy\,\mu s}$ and only 1.5\% of the total flux is below $0.15{\rm\;kJy\,\mu s}$, while for the IP, $A=1.6{\rm\;s^{-1}}$ and $\langle E\rangle=0.24{\rm\;kJy\,\mu s}$, and 13\% of the flux is below.

We can verify whether our MP GPs account for most of the flux by calculating pulse profiles with and without removing pulse rotations where GPs are detected.
As can be seen in Figure~\ref{fig:ek036_regular_emission}, significant flux remains in both MP and IP.
For the MP, even though the remaining signal is brighter in epochs B and D, the fraction is lower: about 18\% in B and D, in comparison with 23\% in A and C.
This again can be understood if the larger detection rate is due to an overall magnification: a larger fraction of the pulses -- and hence of the total flux -- is detected.

Our result is similar (but more constraining) than that of \citet{Majid2011}, who showed that at least $54\%$ of overall pulsed energy flux for the Crab Pulsar is emitted in the form of GPs.
But it is in contrast for what is seen by \citet{Abbate2020} for PSR J1823$-$3021A, where the detected GPs make up only a small fraction of the integrated pulse emission ($4\%$ and $2\%$ for their C1 and C2 components, respectively), and by \citet{Geyer2021} for PSR J0540$-$6919, where the detected GPs only make up $7\%$ of the total flux.
This might indicate a difference in the emission process.
As these authors noted, however, a larger population of undetected GPs may still be hidden below their detection threshold.

For our observations, for both MP and IP, the residual flux is much larger than expected based on the log-normal distribution, thus indicating that the true fluence distribution has more pulses at low fluence (many more for the IP); if additional pulses were emitted also in rotations that we do not detect them, their typical fluence would be the residual flux integrated over one cycle, which is $\sim\!25{\rm\;Jy\,\mu s}$ for MP and a little less for IP.
This is well below our detection limit, so consistent in that sense, but from the distributions shown in Figure~\ref{fig:ek036_energydistribution}, one would expect a much smaller rate than once per pulse period at $25{\rm\;Jy\,\mu s}$.
This might suggest that there are even more but typically fainter bursts (note that it cannot be fainter bursts accompanying the GPs we already detect, since we excluded the full rotations in calculating the residual emission), or that there is some steady underlying emission.
It would be worthwhile to test this with more sensitive future observations.

\subsection{Pulse Phases\label{subsec:phase_distributions}}

\begin{figure*}[ht!]
  \centering
  \includegraphics[width=0.985\textwidth,trim=0 0 0 0,clip]{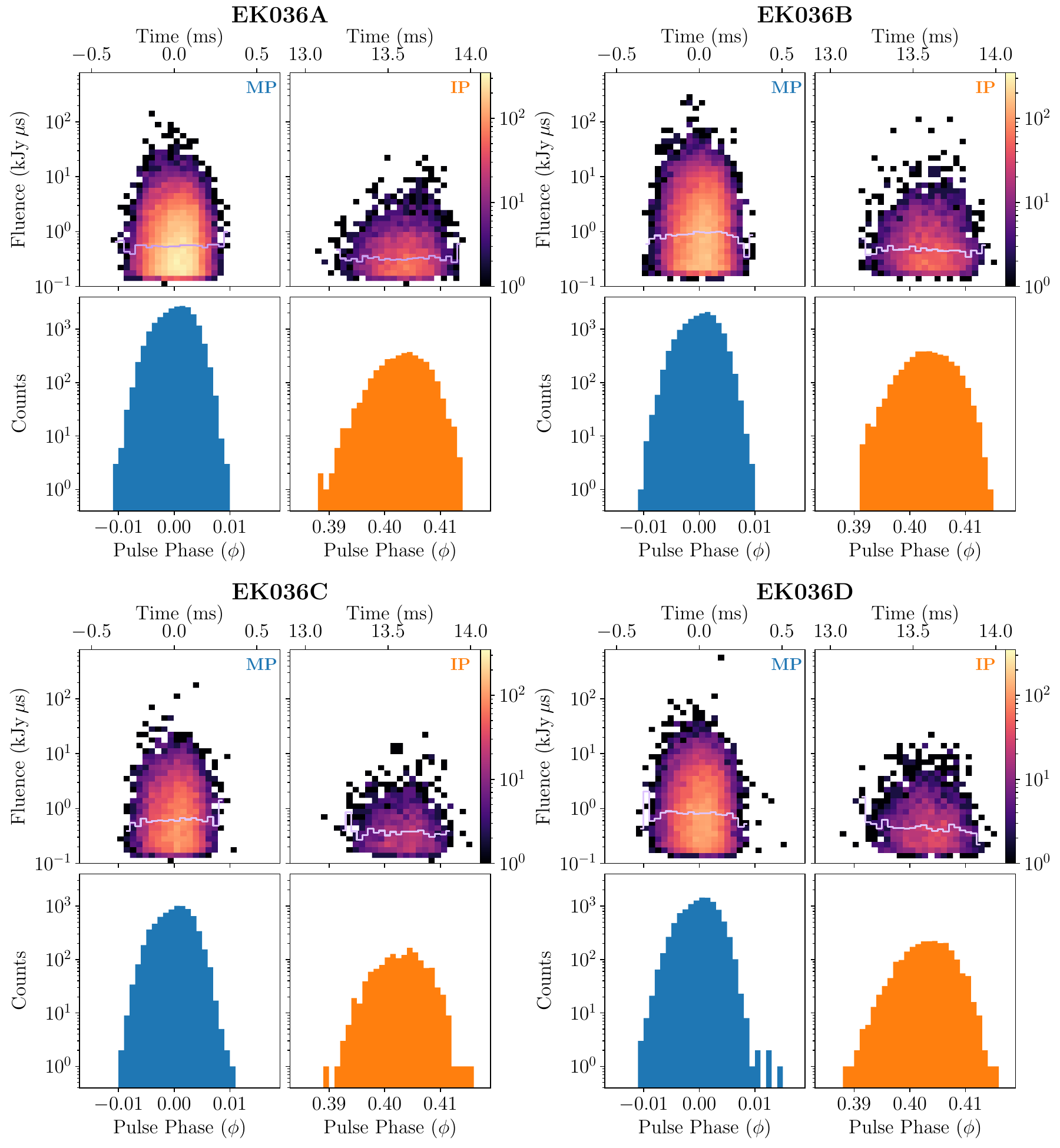}
  \caption{
    MP GP and IP GP fluence and count distributions as a function of pulse phase for each EK036 observation.
    We used pulse phase bins of $0.1\%$ and fluence bins of $0.1{\rm\;dex}$.
    The light purple line in the fluence panels show the median for bins with more than $2$ detected pulses.
  }\label{fig:ek036_phase}
  \vspace{4mm}
\end{figure*}

Defining the time of arrival of a GP as the time when an increase in flux is first detected, the longitude windows where MP and IP GPs occur have total widths of $\sim\!680{\rm\;\mu s}$ and $860{\rm\;\mu s}$ (or $\sim\!7\fdg3$ and $\sim\!9\fdg2$), respectively (averaged over the four epoch).
As can be seen in Figure~\ref{fig:ek036_phase}, the majority of GPs occur within much narrower windows: the root-mean-square deviations around the mean arrival phases are $\sim\!100{\rm\;\mu s}$ and $\sim\!130{\rm\;\mu s}$ (or $\sim\!1\fdg1$ and $\sim\!1\fdg4$), respectively.
The number distribution is roughly Gaussian, with a slightly negative skewness (i.e., a longer tail toward earlier phases and thus with a mode towards later phases).
This was also observed by \citet{Majid2011} at a similar frequency of $1664{\rm\;MHz}$.
In EK036~D, a few MP pulses are detected beyond the range found in the other epochs.
As we will discuss in Section~\ref{subsec:scattering}, these ``outlier'' detections are due to echoes (hence, they are are omitted in our determinations of widths above).

In Figure~\ref{fig:ek036_phase}, we also show the flux distributions as a function of pulse phase, including the median flux of the GPs detected in any given phase bin.
One sees no obvious variation, i.e., no hint of, e.g., brighter pulses having an intrinsically narrower phase distribution.
This suggests that only the probability of seeing a pulse depends on pulse phase.
In our earlier work on these data, where we studied how the pulse spectra and their correlations are affected by scattering \citep{Lin2023}, we concluded that we resolved the regions from which the nanoshots that comprise individual GPs are emitted, and that this is most easily understood if the emitting plasma is ejected highly relativistically, with $\gamma\simeq10^4$ (as was already suggested by \citealt{Bij2021}).
If so, the emission would be beamed to angles much smaller than the width of the phase windows, and the range of phases over which we observe GPs would reflect the range of angles over which plasma is ejected.

\subsection{Arrival Times\label{subsec:arrival_times}}

\begin{deluxetable}{cccccc@{~~~}ccc}
\tabletypesize{\small}
\setlength\tabcolsep{3.8pt}
\tablecaption{Number of Rotations with Multiple Bursts.\label{table:bursts}}
\tablenum{3}
\tablehead{\colhead{Observation}&
  \multicolumn{5}{c}{\dotfill MP\dotfill~~}&
  \multicolumn{3}{c}{\dotfill IP\dotfill}\\[-.7em]
  \colhead{Code}&
  \colhead{2}&
  \colhead{3}&
  \colhead{4}&
  \colhead{5}&
  \colhead{6~~~}&
  \colhead{2}&
  \colhead{3}&
  \colhead{4}}
\startdata
EK036 A & 1820(599) & 200(12) & 24 & 0 & 0 & 144(17) &\phn4 & 2\\
EK036 B & 1431(611) & 170(18) & 22 & 3 & 1 & 237(43) & 16 & 2\\
EK036 C &\phn611(213) &\phn67(\phn4) &\phn6 & 0 & 0 &\phn54(\phn7) &\phn4 & 0\\
EK036 D &\phn934(395) & 117(10) & 23 & 6 & 1 & 116(19) &\phn9 & 0\\
\enddata
\tablecomments{
Numbers in parentheses are those expected if bursts occur randomly; for that case, one does not expect to find any rotations with 4 or more MP bursts or 3 or more IP bursts.
Note that our GP detection method does not differentiate between microbursts and echoes, which becomes important for a few very bright pulses in EK036~D, for which echoes were present.
In addition, we are not able to distinguish microbursts that occur very close together in time.
The number of detections differ from \citet{Lin2023} as a different, more robust, search algorithm is implemented here (see Section~\ref{subsec:gpsearch}).
}
\vspace{-8mm}
\end{deluxetable}

\vspace{-4mm}
Several studies (e.g., \citealt{Karuppusamy2010, Majid2011}) have found that GPs in different rotations are not correlated, and that there is no correlation between MP and IP GPs, but that instead the distribution of the time delays between successive GPs follows an exponential distribution, as expected for a Poissonian process.
Within a given cycle, though, multiple correlated microbursts can occur \citep{Sallmen1999, Hankins2007}.

With our high sensitivity, we can investigate this in more detail.
In Table~\ref{table:bursts} we show the number of rotations in which we detect multiple MP or IP bursts (i.e., double, triple etc.), as well as the number expected (listed only where larger than 0)
for the case where all events are independent,
\begin{equation}
  N_{n} = p_{n} N_{r} = \binom{N_{\rm p}}{n} \left(\frac{1}{N_{r}}\right)^{n}\left(1 - \frac{1}{N_{r}}\right)^{N_{\rm p}-n}N_{r},
\end{equation}
where $p_{n}$ is the probability of a given rotation to have $n$ bursts (assuming a binomial distribution), $N_{r}$ is the total number of rotations observed, and $N_{\rm p}$ is the total number of bursts found (and where for numerical values we inserted numbers from Table~\ref{table:log}: $N_{\rm p} = N_{\rm MP}$ or $N_{\rm IP}$ and $N_{r} = t_{\rm exp}/P_{\rm Crab}$, where $P_{\rm Crab}=33.7{\rm\;ms}$ is the rotation period of the pulsar).
One sees that we detect significantly more multiples than expected by chance\footnote{In \cite{Lin2023}, we wrongly concluded the multiples were consistent with arising by chance. Sadly, we used incorrect estimates of $N_{n}$.}, i.e., some of the detected pulses are composed of multiple, causally related microbursts.

In principle, one could estimate the number of independent bursts, $N^{\rm ind}_{\rm p}$, in each epoch by subtracting from $N_{\rm p}$ the excess pulses from Table~\ref{table:bursts}, but this would not be quite correct since the excess would be relative to estimates made using the total number of observed pulses $N_{\rm p}$, not the (lower) number of independent pulses $N^{\rm ind}_{\rm p}$.
One could iterate, but an easier, unbiased estimate of $N^{\rm ind}_{\rm p}$ can be made using the observed fraction of rotations in which we do not see any bursts, which should equal $N_{0}/N_{r} = p_{0} = (1 - 1 / N_{r})^{N^{\rm ind}_{\rm p}}$.
Solving for $N^{\rm ind}_{\rm p}$, we find that $N^{\rm ind}_{\rm p} = fN_{\rm p}$ with fractions $f$ that are consistent between all epochs, at $91.8\pm0.2$ and $95.2\pm0.5$\% for MP and IP, respectively.
Hence, about 8 and 5\% of the detected MP and IP pulses, respectively, are extra components. Or, as fractions of independent MP and IP pulses, $(6, 1, 0.12)$ and $(4, 0.3, 0.0)\%$, respectively, are causally related double, triple, or quadruple microbursts.

\begin{figure*}[t!]
    \centering
    \includegraphics[width=0.985\textwidth,trim=0 0 0 0,clip]{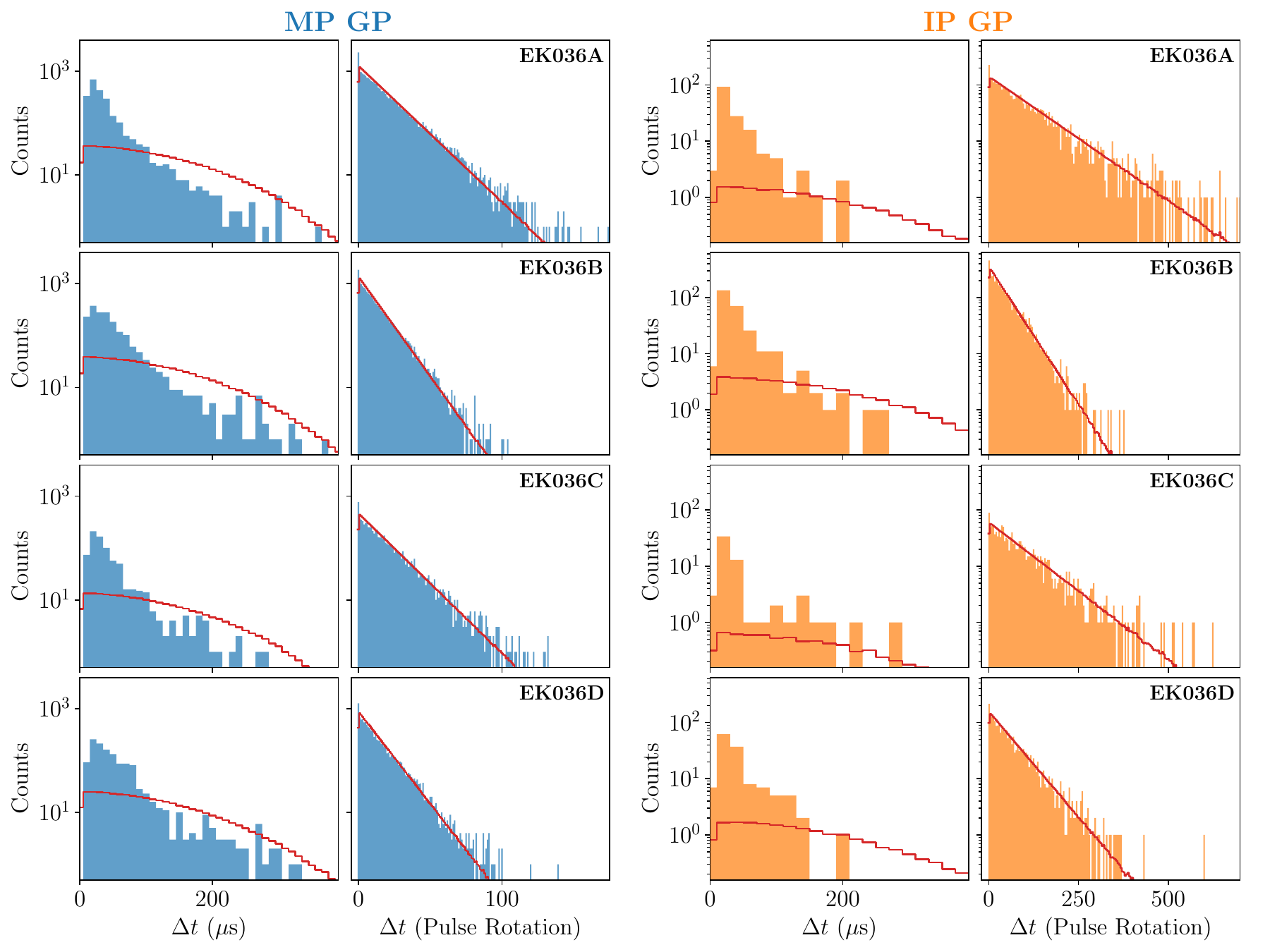}
    \caption{
    Time delays between successive GPs for the MP (in blue) and IP (in orange) components for each EK036 observation.
    On the left MP and IP columns, time delays within a pulse rotation are shown with bins of $10{\rm\;\mu s}$ and $20{\rm\;\mu s}$ for the MP and IP respectively; the low counts in the first bin reflect the minimum separation of $8.75{\rm\;\mu s}$ between detected pulses.
    On the right MP and IP columns, time delays in pulse rotations are shown with bins of $1$ rotation and $4$ rotations for the MP and IP respectively.
    The red lines show the average time delay histograms for $1000$ bootstrap iterations, in which we randomized the rotation in which a pulse was seen (but not the phase, to keep the observed phase distribution).
    }\label{fig:ek036_dt}
    \vspace{4mm}
\end{figure*}

To investigate the distributions further, we show histograms of the time delay between pulses in Figure~\ref{fig:ek036_dt}.
Overdrawn are expectations for randomly arriving, independent pulses.
We constructed these by bootstrapping, where we repeatedly reassign new random pulse cycles to our observed sets of pulses, and then recalculate the time delay distributions.
Note that in our bootstraps, we do not randomize pulse phase, so that the observed phase distribution is correctly reflected in the time delays.
One sees that as a function of pulse cycle (right column panels for MP and IP GPs in Figure~\ref{fig:ek036_dt}), the observed histograms follow the expected exponential distribution (although the observed counts are slightly lower than the expected ones because not all pulses are independent, as is implicitly assumed in the bootstraps).

For the time delays between pulses that occur in the same cycle (left column panels for MP and IP GPs in Figure~\ref{fig:ek036_dt}), the observed distributions are very different from those expected for randomly occurring bursts.
One sees a large peak at short delays, representing the excess microbursts from Table~\ref{table:bursts}, following a roughly exponential distribution with a mean time between bursts of $\sim\!30{\rm\;\mu s}$ or so.
Intriguingly, at somewhat larger time difference, there seem to be fewer bursts than expected for independent events.
This suggests that while a given detection has an enhanced probability of being in a group of causally related microbursts, the occurrence of a burst also suppresses the likelihood of another, independent, burst being produced in the same rotation.
Thus, our results confirm that GPs are often composed of multiple microbursts, and they indicate that another, independent GP is less likely to occur right after.

\begin{figure*}[ht!]
    \centering
    \includegraphics[width=0.985\textwidth,trim=0 0 0 0,clip]{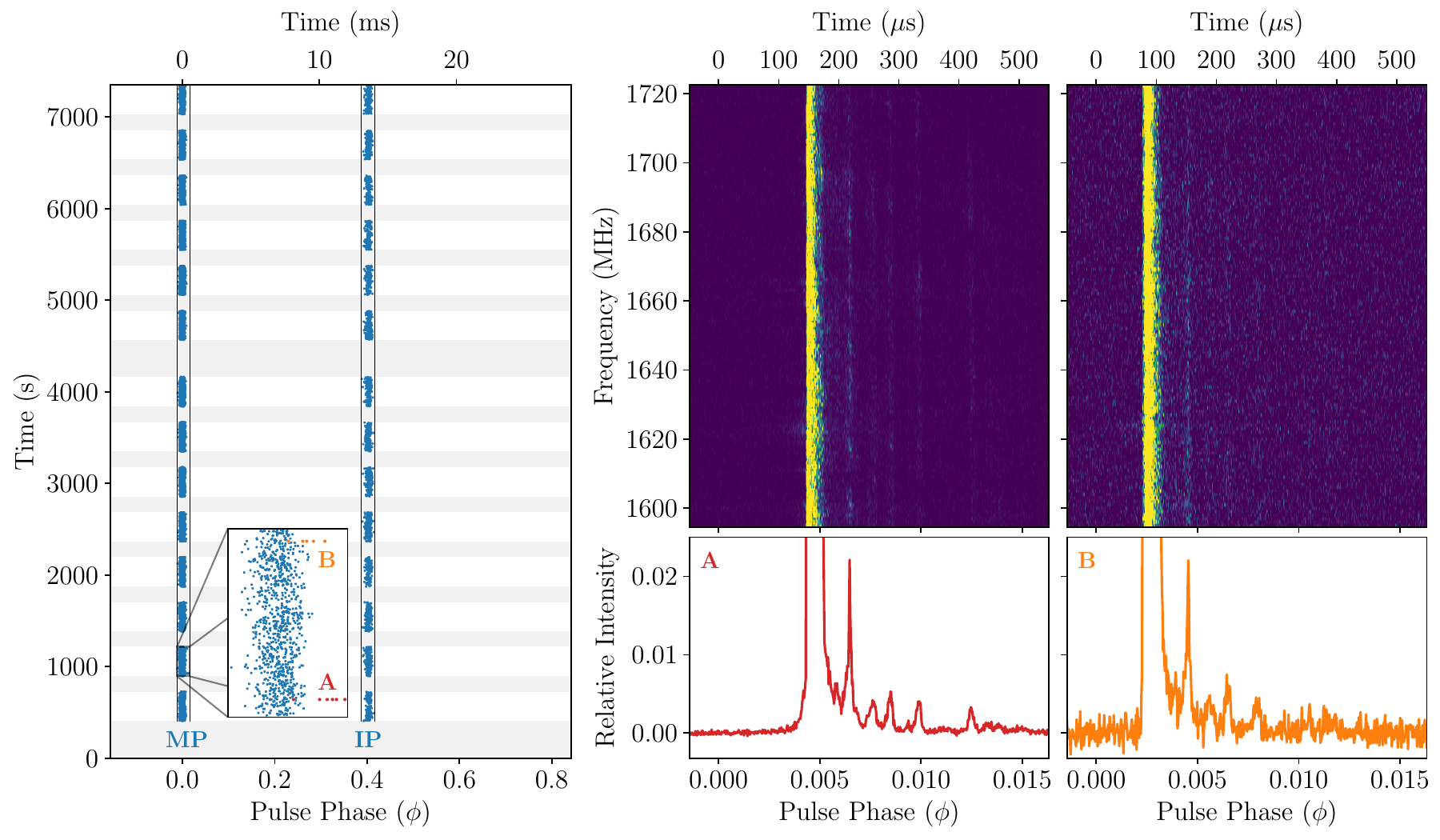}
    \caption{\emph{Left\/}: MP GPs and IP GPs detected in the EK036~D data.
    The gray shaded regions indicate when the telescope was not observing the Crab Pulsar and the black vertical lines mark our MP GP and IP GP windows.
    In the inset, we show two pulse rotations containing the brightest GPs ``A'' and ``B'', in red and orange respectively.
    \emph{Right, Top\/}: Waterfalls of the two brightest pulses in EK036~D with $1{\rm\;\mu s}$ time resolution and $1{\rm\;MHz}$ frequency resolution.
    \emph{Right, Bottom\/}: Pulse profile of the two brightest pulses in EK036~D with $1{\rm\;\mu s}$ time resolution scaled to the peak of each pulse.
    Pulses ``A'' and ``B'' show similar features and we conclude that during the EK036~D observations, weak echoes were present at large delays.}
    \label{fig:ek036d}
    \vspace{4mm}
\end{figure*}

\begin{figure*}[ht!]
    \centering
    \includegraphics[width=0.985\textwidth,trim=0 0 0 0,clip]{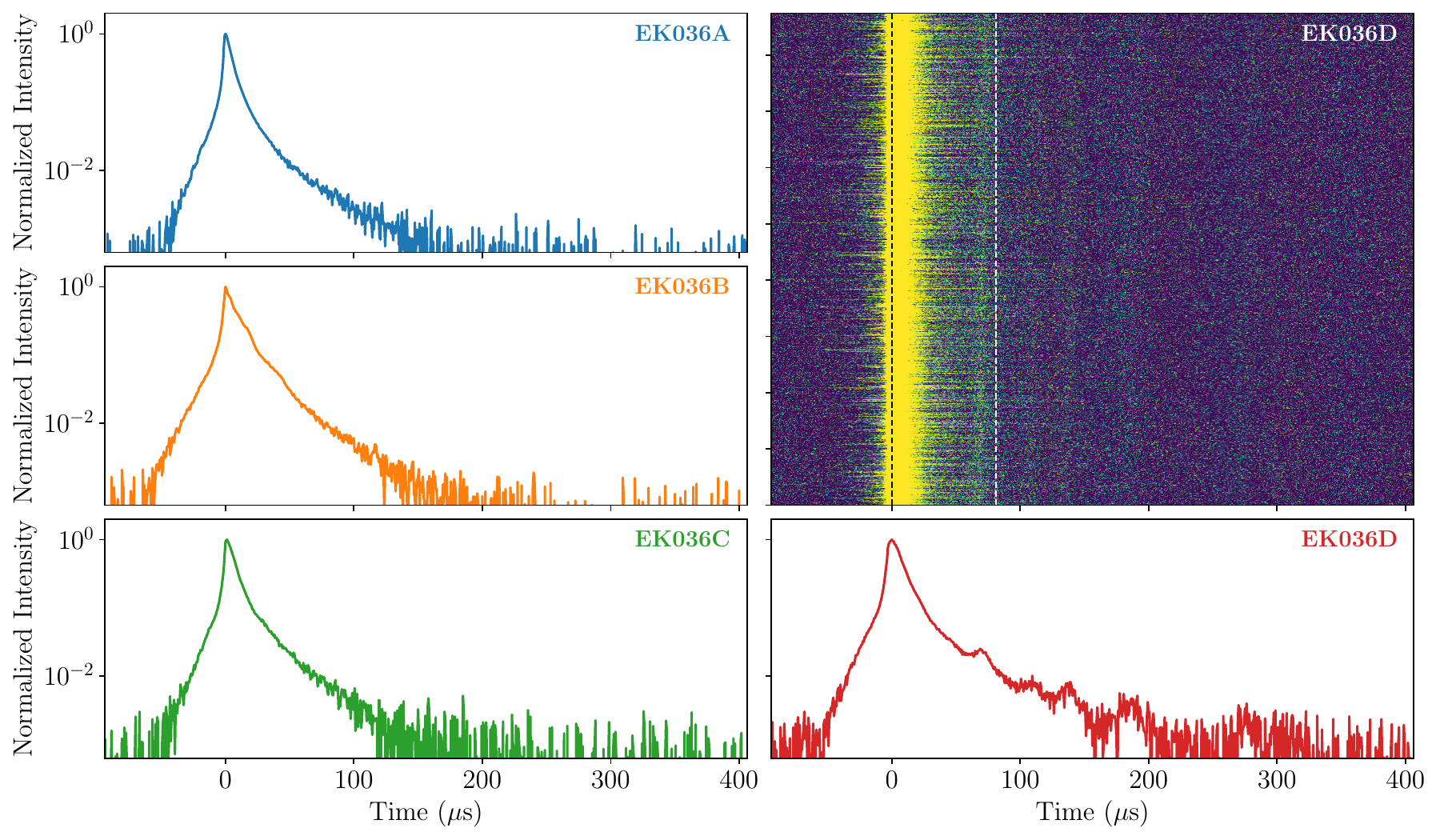}
    \caption{
    \emph{Line plots\/}: SVD approximation of the MP pulse profile for all observations.
    In EK036~B, echoes are seen close to the profile's peak (see \citealt{Lin2023} for more details).
    The profile for EK036~D shows multiple weak echoes up to $\sim\!300{\rm\;\mu s}$.
    \emph{Image\/}: The MP pulse stack for EK036~D, using a logarithmic colour scale to bring out faint features.
    Each pulse is aligned by correlating with the rotation with the brightest pulse in EK036~D (which is appears to be a simple single microburst) and then normalized by the intensity at time $0$ (the black dashed line).
    The echoes appear to move out over time, as one can see by comparing the location of the most prominent faint echo with the dashed white vertical line near it (time is increasing both upwards and to the right in this image).
    }
    \label{fig:ek036_svd}
    \vspace{4mm}
\end{figure*}

\subsection{Scattering Features}\label{subsec:scattering}

In Figure~\ref{fig:ek036_phase}, one sees that in EK036~D, several MP GPs were detected at pulse phases quite far from the median phase.
To investigate this, we looked at the arrival times of all GPs detected in EK036~D (see left panel of Figure~\ref{fig:ek036d}).
We found that the outliers occurred in two pulse rotations, which turned out to contain the brightest GPs in EK036~D.
Looking at the pulse profiles of these brightest GPs, one sees that they are very similar (see right panels of Figure~\ref{fig:ek036d}).
In fact, closer examination reveals that all of the brightest GPs detected in EK036~D show similar pulse profiles.
This implies that the pulses far from the median pulse phase arrive late because they are actually weak echoes of the main burst, with amplitudes down to $\sim\!0.4\%$ of the peak flux and delays up to $\sim300{\rm\;\mu s}$.

In Figure~\ref{fig:ek036_svd}, we show singular value decomposition (SVD) approximations of the average MP GP profile for each epoch (for the IP, too few bright pulses were available).
This was created from MP GP rotations with peak intensities greater than $200{\rm\;Jy}$ and seemingly single peaks, aligned using time offsets found by correlation with a reference pulse.
To avoid giving too much weight to the brightest pulses, and thus risking that remaining substructure enters the average profile, we normalized each rotation by the intensity at the correlation maximum before doing the SVD.
One sees that all profiles are fairly sharply peaked, but sit on top of a base, which has the expected asymmetric part extending to later time due to scattering, as well as a more symmetric component, likely resulting from the collective effect of faint microbursts.
Comparing the epochs, one sees that for EK036~A-C, the profile dropoff is relatively smooth and becomes undetectable after $\sim\!200{\rm\;\mu s}$, while in EK036~D, the tail is much longer, extending to $\sim\!400{\rm\;\mu s}$, and is much more bumpy.

Almost certainly, all bumps are echoes, including those at shorter delay in EK036~B (more clearly seen in the linear-scale plots in \citealt{Lin2023}),
Indeed, looking carefully at the stack of profiles in Figure~\ref{fig:ek036_svd}, one sees that the echoes in EK036~D drift in time, moving slightly further away from the MP during the observation, with perhaps even a hint that echoes further away from the main bursts drift faster than those closer in.
(Note that this stack is not completely linear in time, although given that the GP detection rate is roughly constant throughout, it is not far off.)
This change in time is expected for echoes off a structure with changing distance from the line of sight, and indeed has been seen for a very prominent echo by \cite{Backer2000, Lyne2001}.
Overall, our observations suggests echoes are common, as also concluded from daily monitoring at $600{\rm\;MHz}$ by Serafin-Nadeau et al.\ (2023, in prep.).

\newpage
\section{Summary of Conclusions} \label{sec:conclusion}

The fine time resolution and high sensitivity in our beamformed EVN data allowed us to confidently detect $65951$ GPs with fluences above $\sim150{\rm\;Jy\,\mu s}$ over a short period of $7.32{\rm hr}$.
Within each of our four observations, we found that the GP detection rates are fairly constant, but that between epochs they differ by a factor of $\sim\!2$.
Similar changes were seen previously, and were suggested by \cite{Lundgren1995} to reflect changes in overall magnification of the scattering screens along the line of sight.

The changes in magnification are consistent with the pulse fluence distributions, which are power-law like at high fluence, but with a flattening at lower fluences; the distributions from the different epochs can be shifted to each other with a change in fluence scale.
We noted that the fluence distributions are similar to what is expected for log-normal distributions, but found that the residual signals seen in the GP phase windows after removing the GPs we detected were larger than expected if the log-normal distribution continued also below our detection limit.
Nevertheless, it suggests that with only somewhat more sensitive observations, it should be possible to get a fairly complete sampling of all GPs that contribute to the average flux, at least for the MP component.

Analyzing the pulse phase distributions, we confirm previous observations showing that the majority of GPs occur within very narrow phase windows.
Furthermore, we observe no significant variations in the median flux distributions as a function of pulse phase.
This suggests that it is the probability of observing a pulse that depends on pulse phase, not its energy, implying that the angle within which a pulse is emitted is much narrower than the rotational phase window, as expected if the plasma causing them is travelling highly relativistically \citep{Bij2021, Lin2023}.

With our high detection rates, we were able to investigate the distribution of time delays between successive bursts within the same pulse rotation.
We detect a larger number than expected if all bursts were due to a Poissonian process, and infer that $\sim\!5\%$ of bursts come in groups of 2 or 3 causally related microbursts, with a typical separation in time of $\sim\!30{\rm\;\mu s}$.

Additionally, our high sensitivity revealed weak echo features for individual bright pulses, which drift slightly but significantly even over our timescales of just a few hours.
We infer that echo events are not rare.

Given our findings, we believe even more sensitive follow-up studies of the Crab Pulsar would be very useful.
This would be possible using more small dishes (spaced sufficiently far apart that the Crab Nebula is well-resolved) and by recording a larger bandwidth.

\section*{Acknowledgements}
We thank the anonymous referee for their comments, which improved the clarity of this manuscript.
We thank the Toronto Scintillometry group, and in particular Nikhil Mahajan, for useful discussion on GP statistics.
Computations were performed on the Niagara supercomputer at the SciNet HPC Consortium \citep{Loken2010, Ponce2019}.
SciNet is funded by: the Canada Foundation for Innovation; the Government of Ontario; Ontario Research Fund - Research Excellence; and the University of Toronto.
M.Hv.K. is supported by the Natural Sciences and Engineering Research Council of Canada (NSERC) via discovery and accelerator grants, and by a Killam Fellowship.

\facilities{The European VLBI Network (EVN) is a joint facility of independent European, African, Asian, and North American radio astronomy institutes.
Scientific results from data presented in this publication are derived from the following EVN project codes: EK036~A-D.}

\software{
  astropy \citep{AstropyCollaboration2013, AstropyCollaboration2018, AstropyCollaboration2022},
  Baseband \citep{VanKerkwijk2020},
  CALC10 \citep{Ryan1980},
  numpy \citep{Harris2020},
  matplotlib \citep{Hunter2007},
  pulsarbat \citep{Mahajan2022},
  scipy \citep{Scipy2020},
  tempo2 \citep{Hobbs2012}.}

\bibliographystyle{aasjournal}
\bibliography{main}

\end{document}